\documentclass[preprint,prd,showkeys,showpacs]{revtex4}
\pdfoutput=1
\usepackage{amsfonts,amsmath,amssymb}
\usepackage{enumerate}
\usepackage{hyperref}
\usepackage{graphicx}
\newcommand{\be}{\begin{equation}}
\newcommand{\ee}{\end{equation}}

\newcommand{\bea}{\begin{eqnarray}}
\newcommand{\eea}{\end{eqnarray}}

\newcommand{\bal}{\begin{align}}
\newcommand{\eal}{\end{align}}

\renewcommand{\d}{{\rm d}}

\newcommand{\bes}{\begin{subequations}}
\newcommand{\ees}{\end{subequations}}

\newcommand{\nn}{\nonumber}

\newcommand{\ra}{\rightarrow}
\newcommand{\cN}{{\cal N}}

\begin{document}

\title{Comments on $AdS_2$ solutions from M2-branes on complex curves and the backreacted K\"ahler geometry}
\author{Nakwoo Kim}
\email{nkim@khu.ac.kr}
\affiliation{Department of Physics
and Research Institute of Basic Science, \\ Kyung Hee University,
Seoul 130-701, Korea }



\begin{abstract}
We consider $AdS_2$ solutions of M-theory which are obtained by 
twisted compactifications of M2-branes on a complex curve. They are of 
a generalized class,  in the sense that the non-abelian part of the connection for
the holomorphic bundle over the supersymmetric cycle is nontrivial. 
They are solutions of $U(1)^4$ gauged supergravity in $D=4$, with magnetic flux over
the curve, and then uplifted to $D=11$. We discuss
the behavior of conformal fixed points as a function of the non-abelian connection.
We also describe how they fit into the general description of wrapped M2-brane
$AdS_2$ solutions and their higher-order generalizations, by showing that they satisfy the master equation
for the eight-dimensional K\"ahler base space. 
\end{abstract}


\maketitle

\section{Introduction}
Since the proposal of AdS/CFT correspondence \cite{Maldacena:1997re}, 
we have witnessed copious examples of $AdS$ solutions
in String/M-theory which are all potentially dual to strongly interacting conformal field theories
in various dimensions. The basic examples are of course flat M2, D3 and M5 branes which guarantee
maximal supersymmetries in D=3, 4, 6. Notably, these branes may provide lower-dimensional,
less supersymmetric $AdS$ solutions when partially wrapped on supersymmetric cycles. On the field
theory side, this procedure corresponds to topologically 
twisting the theory \cite{Witten:1988ze} through coupling to a combination of R-symmetry currents.
The prescription for such configurations in lower-dimensional gauge supergravity theories was
first illustrated in \cite{Maldacena:2000mw}. This work considered branes wrapped on
two-cycles in Calabi-Yau manifold, and the generalization to higher-dimensional cycles was given 
in e.g. \cite{Acharya:2000mu}\cite{Gauntlett:2000ng}. Schematically, one turns on the magnetic fields
for bulk gauge fields which encode the twisting of the metric in higher dimensions, so that 
the effect of curvature on the supersymmetric cycle is cancelled. In particular, for K\"ahler 2-cycles
this means we assign precisely the spin-connection of the two-cycle to the diagonal 
$U(1)$ part of the bulk gauge field. Then the supersymmetry is retained, albeit the curvature on the 
world-volume of the brane alone would 
not allow constant spinor. Once the branes and supersymmetric
cycles are identified, at least in the original prescription
\cite{Maldacena:2000mw}\cite{Acharya:2000mu}\cite{Gauntlett:2000ng}\cite{Gauntlett:2001jj}\cite{Gauntlett:2002rv}
one obtains unique BPS equations for each wrapped brane configuration when the cycles are hyperbolic space.

For such magnetic brane solutions relevant to 2-cycles, more general solutions were constructed
with multiple non-vanishing $U(1)$ charges, where the sum of magnetic fields
 still exactly cancel the spin connection part
in the Killing equations \cite{Cucu:2003bm}\cite{Cucu:2003yk}\cite{Cacciatori:2009iz}. 
Their interpretation as
general wrapped branes and the description of dual conformal field theory are given in \cite{Bah:2011vv}\cite{Bah:2012dg}. For concreteness,
let us take a two-cycle in Calabi-Yau three manifold (CY3). 
For a holomorphic curve as K\"ahler two-cycles in CY3, locally the geometry is a
holomorphic $\mathbb{C}^2$ bundle over the curve, so in general the connection takes values in $U(2)$. 
The supersymmetry constrains the diagonal $U(1)$ part
of the $U(2)$, leaving the $SU(2)$ part to one's disposal. It is pointed
out in \cite{Bah:2012dg} that 
choosing the $SU(2)$ connection appropriately it should be possible to find
a solution interpolating CY3 and CY2 as the embedding space. 
In gauged supergravity the implementation is quite simple:
originally for two-cycles in CY3 we take the diagonal $U(1)$ among $SO(4)$ gauge fields (which
of course represent the isometry of $\mathbb{C}^2=\mathbb{R}^4$ bundle over the holomorphic curve) and assign it 
the same value as the spin connection of the two-cycle. For the 
generalization we take $U(1)\times U(1)
\subset SO(4)$, and demand the {\it sum} of two $U(1)$ connections be the same as the spin connection 
of the two-cycle. This prescription leads to a one-parameter generalization of BPS equations for M5-branes
wrapping a two-cycle in CY3 \cite{Bah:2011vv}\cite{Bah:2012dg}, and eventually an infinite number of
new ${\cal N}=1$ $AdS_5/CFT_4$ examples. The computation of central charges on both sides match perfectly. 
This is generalized to D3 branes and also to different supersymmetric cycles in \cite{Benini:2012cz}\cite{Benini:2013cda}\cite{Karndumri:2013iqa}. 

We note that, a very similar idea can be applied also to the case of flat 2-cycles, leading
to $AdS\times T^2$ or $AdS\times \mathbb{R}^2$ solutions \cite{Almuhairi:2010rb}\cite{Almuhairi:2011ws}. 
These were dubbed magnetic brane
solutions and studied further in \cite{Donos:2011pn}\cite{Almheiri:2011cb}\cite{Lu:2013eoa}.
More generally, black holes of $D=4$ gauged supergravity with the geometry of $AdS_2\times \Sigma_g$, with  $\Sigma_g=S^2,H^2,T^2$ have been 
extensively studied in \cite{Halmagyi:2013sla}\cite{Halmagyi:2013qoa}.

In this article we consider M2-branes wrapped on $H^2$, especially the generalization with 
nonabelian part of connection on $H^2$. The relevant $D=4$ gauged supergravity has $U(1)^4$ gauge fields and is a consistent truncation of maximal $SO(8)$ gauged supergravity theory.
One may consider in this case two-cycles in CY2, CY3, CY4, and CY5 manifolds. For each of
them in the original construction \cite{Gauntlett:2001qs} we turn on one, two, three, and four
of $U(1)$'s respectively to cancel out the effect of spin connection on the two-cycle. 
Then it turns out \cite{Gauntlett:2001qs}, there exist $AdS_2$ fixed points
for CY4 and CY5. The generalization mentioned above leads to a one-parameter family of
$AdS_2$ fixed points interpolating CY4 and CY5. We also describe various holographic renormalization 
group flows between $AdS_4$ and $AdS_2\times H^2$ solutions.

We will also discuss the uplifted $D=11$ solutions of $AdS_2\times H^2$ solutions.
Containing an $AdS_2$ factor and being supersymmetric together imply
 that the solutions can be rewritten in the canonical form, where the internal 
nine-dimensional space is a warped and twisted circle fibration over eight-dimensional K\"ahler space, 
satisfying \cite{Kim:2006qu}
\be
\Box_8 R - \frac{1}{2} R^2 + R_{ij} R^{ij} = 0 . 
\label{ke00}
\ee
It is in principle possible to solve this equation first and then
using the geometric data of the K\"ahler base to construct the whole $D=11$ solution. 
In this paper we start with the uplifted $D=11$ solution and employ a natural ansatz to 
check the above equation is indeed satisfied.

Supersymmetric solutions written in canonical form can provide useful information more readily. 
The supersymmetry relations are derived from the existence of Killing spinors and also by exploiting
the properties of all spinor bi-linears. As such, solutions in canonical form immediately give full
information on the Killing spinors. This can be very useful for instance when one looks for  supersymmetric 
brane embeddings in a given supersymmetric background. 
References which successfully applied this technique to 
nontrivial backgrounds can be found e.g. in \cite{Bah:2013wda}\cite{Gauntlett:2006ai}.

In the next section we give the action, the BPS equations and solve for the fixed points. 
In Sec. 3 we discuss the uplift to $D=11$, and as a consistency check we verify 
that the nine-dimensional internal space satisfies the general condition given in \cite{Kim:2006qu}
and discuss generalization to other dimensions. We conclude in Sec. 4. 

\section{$AdS_2$ solutions}
In this section we present multiply-charged magnetic brane solutions in $D=4$ gauged supergravity. As far 
as we know such solutions are first systematically constructed in \cite{Cucu:2003yk}, especially 
in Section 8.

The supergravity model we employ here is the $U(1)^4$ truncated action of 
$\cN=8, \, D=4$ gauged supergravity \cite{Duff:1999gh}\cite{Cvetic:1999xp}. 
After truncation, the bosonic field content reduces to graviton, four 
vector fields $A_\alpha \, (\alpha=1,2,3,4)$, and three real scalar fields $\vec \phi$. The action
is 
\be
{\cal L} = \frac{1}{2\kappa^2} \sqrt{-g} \left[
R - \frac{1}{2} (\partial \vec \phi)^2 - 2 \sum_\alpha e^{\vec a_\alpha \cdot \vec \phi} F^2_\alpha 
 - V \right] , 
\label{action}
\ee
where
\be
V = - \frac{4}{L^2}( \cosh \phi_{12} +\cosh \phi_{13} +\cosh \phi_{14} ) . 
\ee
$L$ is a constant which sets the curvature of the $AdS_4$ vacuum solution.
The three scalars $\vec \phi=(\phi_{12},\phi_{13},\phi_{14})$ 
can be alternatively expressed in terms of $X_\alpha$ in the following way. 
\be
X_\alpha = {\rm exp} ( - \vec a_\alpha \cdot \vec \phi /2 ) , 
\ee
where 
\begin{equation}
\begin{array}{rclrrcl}
\vec a_1  &=& (1,1,1) , \, & \vec a_2 &=& (1,-1,-1) , \cr
 \vec a_3 &= &(-1,1,-1), \, &\vec a_4  &=& 
(-1,-1,1).
\end{array}
\end{equation}
Note that $\prod_{\alpha=1}^4 X_\alpha = 1 $. 

It is well known that the maximal gauged supergravity in $D=4$, and in particular its $U(1)^4$ truncated version \eqref{action}, is a consistent truncation of $D=11$ supergravity. It means 
that any solution of \eqref{action} we may find leads to an exact solution in $D=11$. 
The uplifting formula for the metric tensor is
\be
\d s^2 = \Delta^{2/3} \d s^2_4 +
{2L^2} \Delta^{-1/3} \sum_\alpha X^{-1}_\alpha
( \d\mu^2_\alpha + \mu^2_\alpha (\d\phi_\alpha + {2}A_\alpha/L)^2 ) , 
\label{uplift}
\ee
where $\mu_a$ are angular variables parametrizing $S^4$, i.e. $\sum^4_{\alpha=1} \mu^2_\alpha = 1$. 
The warp factor $\Delta = \sum X^{}_\alpha \mu^2_\alpha$. 

As discussed above, our generalized ansatz interpolates the two $AdS_2$ solutions of 
M2-branes wrapped on 2-cycles in Calabi-Yau four-folds and five-folds. For the gauge fields,
\be
F_1 = F_2 = F_3 = -\frac{z lL}{2} \, {\rm vol}(\Sigma) , \quad 
F_4 = -\frac{(1-3z) lL}{2}  {\rm vol}(\Sigma) .
\ee
Here we may choose the 2-cycle $\Sigma$ to have constant curvature, like $S^2,T^2$ and $H^2$.
$l$ is the scalar curvature of $\Sigma$ and without losing generality we may 
scale it to be $1,-1$ when it is nonzero. For $T^2$, we may first rewrite $l \ra l/z$ before sending
$l$ to zero. In particular, for $T^2$ we have in general 
\be
F_1 = F_2 = F_3 = -\frac{zL}{2} \, {\rm vol}(\Sigma) , \quad 
F_4 = \frac{3zL}{2}  {\rm vol}(\Sigma) .
\ee

Now for the rest of the bosonic fields, 
in order to respect the same symmetry we constrain the scalar fields as follows,
\be
X_1= X_2 = X_3 = e^{-\phi/2} , \quad X_4 = e^{3\phi/2} . 
\ee
We note here that\ $z=1/3$ corresponds to the Calabi-Yau four-fold case, while 
$z=1/4$ is for the Calabi-Yau five-fold solution in \cite{Gauntlett:2001qs}. For the metric our ansatz
is 
\be
\d s^2_4 = e^{2f} (-\d t^2 + \d r^2) + e^{2g} \d s^2(\Sigma) . 
\ee

The supersymmetry transformation rules for the $U(1)^4$ model can be found for instance
in \cite{Gauntlett:2001qs}\cite{Donos:2011pn}. There are four spinor parameters $\epsilon^{\alpha} \, 
(\alpha=1,2,3,4)$ and just like the 1/8-BPS condition in \cite{Gauntlett:2001qs}, we choose 
$\epsilon^\alpha=0$ for $\alpha=2,3,4$. For generic values of $z$ the BPS equations are given as 
\begin{align}
e^{-f} f' &= - \frac{1}{2\sqrt{2}L} ( 3 e^{-\phi/2} + e^{3\phi/2} ) + \frac{l e^{-2g}
L}{2\sqrt{2}} ( 3z e^{\phi/2} + (1-3z) e^{-3\phi/2}) , 
\\
e^{-f} g' &= - \frac{1}{2\sqrt{2}L} ( 3 e^{-\phi/2} + e^{3\phi/2} ) - \frac{l e^{-2g}
L}{2\sqrt{2}} ( 3z e^{\phi/2} + (1-3z) e^{-3\phi/2}) , 
\\
e^{-f} \phi' &= - \frac{1}{\sqrt{2}L} ( e^{-\phi/2} - e^{3\phi/2} ) + \frac{l e^{-2g}
L}{\sqrt{2}} (z e^{\phi/2} - (1-3z) e^{-3\phi/2}) . 
\end{align}
One can readily check that for $z=1/3$ our equations are the same as (3.19) of \cite{Gauntlett:2001qs} (up to redefinition $L=1/e$ in the notation of \cite{Gauntlett:2001qs}). On the other hand, for $z=1/4$ an obvious solution is $\phi=0$ and the rest of the above equations become
(3.23) of \cite{Gauntlett:2001qs}. 

In general the above equations determine the holographic RG flow obtained by twisting the
maximally supersymmetric $AdS_4$ solution. To identify the $AdS_2$ fixed points,
we demand $g'=\phi'=0$, and it is straightforward to deduce
\be
z e^{4\phi} + (1-6z) e^{2\phi} + 1 - 3z = 0 . 
\label{qe1}
\ee
Solving for $e^{2\phi}$, we have 
\be
e^{2\phi} = \frac{6z-1 \pm \sqrt{(12z-1)(4z-1)}}{2z} . 
\label{br}
\ee
This equation is essentially the same as the first equation of Eq.(8.9) in \cite{Cucu:2003yk}. 

Since $e^{2\phi}>0$ by construction, this result provides a constraint for $z$. 
One can easily check that the roots are both positive for $1/4<z<1/3$, i.e.
between the two $AdS_2$ solutions in \cite{Gauntlett:2001qs}.  When $z\ge 1/3$
only one of the solutions give a positive value for $e^{2\phi}$. The result in \eqref{br}
is independent of $l$, but when we evaluate $e^{-2g}$ we find that it is negative if
$l>0$. $l=0$ leads to interesting solutions of type $AdS_2\times \mathbb{R}^2$, 
which was studied in detail in \cite{Donos:2011pn}\cite{Almheiri:2011cb}. 

\begin{figure}[htb]
\centering
\includegraphics[height=5in]{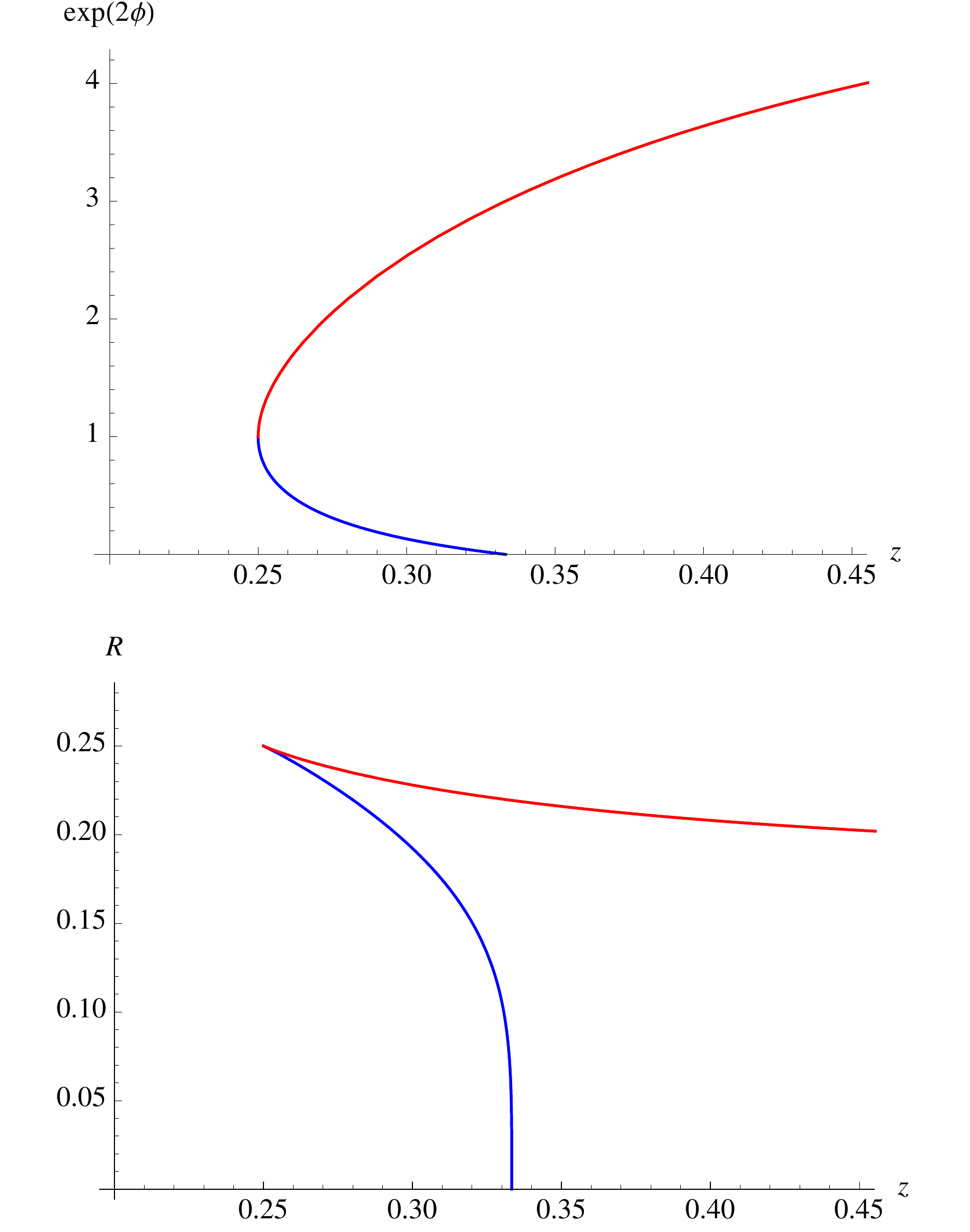}
\caption{The value of $e^{2\phi}$ and $R_{AdS_2}$ vs. parameter $z$ for $AdS_2$ solutions.
The upper (lower) branch is for positive (negative) sign in \eqref{br}.}
\label{fff1}
\end{figure}

\begin{figure}[htb]
\centering
\includegraphics[height=6.5in]{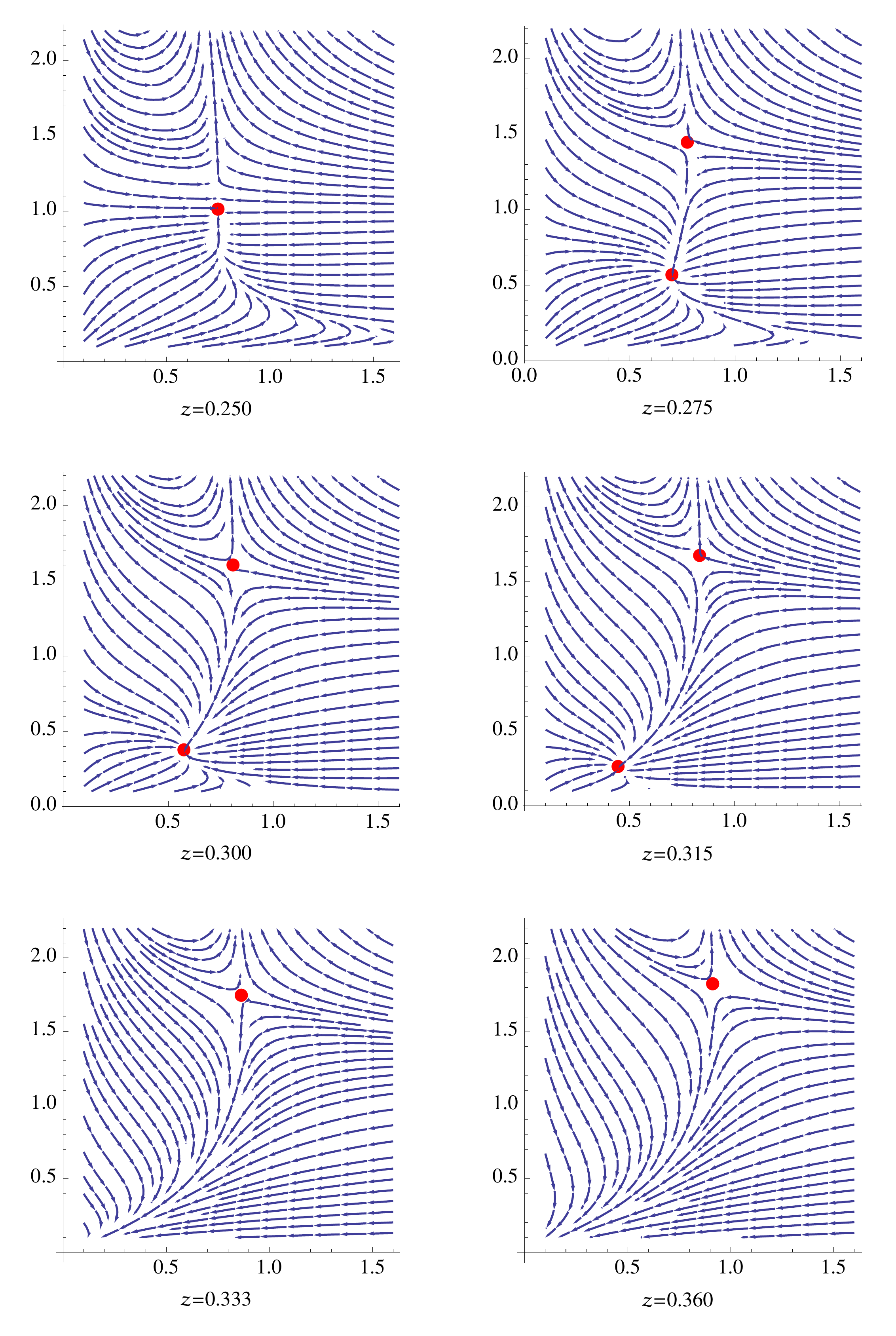}
\caption{Holographic RG flows for different values of $z$. The horizontal (vertical) axis is 
$e^{2g}$ ($e^{2\phi}$). Red dots represent the 
$AdS_2$ fixed point.}
\label{f2}
\end{figure}

In Fig. \ref{fff1} we show the behavior of $e^{2\phi}$ and
$e^{2g}$ as functions of $z$ at $AdS_2$ fixed points. We can also construct the flow solutions starting from the maximally supersymmetric $AdS_4$ 
solutions and going down to $AdS_2$ or other singular configurations. The plots of such 
flow solutions for $z=0.250, 0.275, 0.300, 0.315, 0.333, 0.360$ are given in Fig. \ref{f2}.
One can see how the conformal fixed points ``move" in $(e^{2g}, e^{2\phi})$ space, as we 
smoothly vary the parameter $z$. 

\section{Uplift to $D=11$ and general description}
It is straightforward to construct $D=11$ supergravity solution 
using the uplifting formula given above in \eqref{uplift}. For the metric, we have
\begin{align}
\d s^{2}_{11} &= \Delta^{2/3}
\Big[ R_{AdS_2}^2 
\d s^2(AdS_2) + e^{2g}  \d s^2(H_2)  
\nn \\
&  
+ \frac{e^{-3\phi/2}L^2}{\Delta}
\Big(
e^{2\phi} \sum_{\alpha=1}^3 ( \d\mu^2_\alpha + \mu^2_\alpha D\phi^2_\alpha)
+ \d\mu^2_4 + \mu^2_4 D\phi^2_4
\Big)
\Big]  . 
\label{11sol}
\end{align}

The radius $R$ of $AdS_2$ part in $D=4$ is given by
\be
\frac{1}{R_{AdS_2}} = \frac{3 e^{-\phi/2} + e^{3\phi/2}}{\sqrt{2}L} , 
\ee
Here it is understood that $\phi$ is determined by the BPS condition \eqref{br}. 
We note that the ratio of radii for $AdS_2$ part and $H^2$ part is given simply as
\be
{R_{H^2}^2}/{R_{AdS_2}^2} = 12z - 1 . 
\label{sr}
\ee

For our general ansatz, the warp factor is 
\be
 \Delta = e^{-\phi/2} (\mu^2_1 + \mu^2_2 + \mu^2_3 ) + e^{3\phi/2} \mu^2_4 . 
\ee
And the twists are such that 
\be
\d (D \phi_1 ) =\d(D \phi_2 )=\d(D \phi_3 ) =-z {\rm vol}(H^2), \quad
\d(D \phi_4 ) = (3z-1) {\rm vol}(H^2) . 
\ee

It is known that for a pure M2-brane configuration with unbroken supersymmetry 
and an $AdS_2$ factor, the solution can be always expressed (at least locally) 
in the following form
\cite{Kim:2006qu}\cite{Gauntlett:2007ts}. 
\be 
\d s^2_{11} = e^{-2B/3} \left[ \d s^2 ( AdS_2 )
+ e^{B} \d s^2_8 + (\d u+ P)^2
\right] . 
\label{m2conf}
\ee
Here the eight-dimensional space ${\cal M}_8$ with metric $\d s^2_8$ should be K\"ahler and its Ricci tensor and 
the scalar curvature satisfy
\be
\Box_8 R - \frac{1}{2} R^2 + R_{ij} R^{ij} = 0 .
\label{ke}
\ee
Then the entire $D=11$ supergravity configuration is determined by the data of ${\cal M}_8$. In 
particular, $\d P$ is the Ricci two-form for ${\cal M}_8$, i.e. for the Ricci 2-form ${\cal R}$, 
${\cal R} = \d P$. The four-form gauge field of $D=11$ supergravity is given by 
$G_4 = F \wedge {\rm 
Vol} (AdS_2)$. And 
\begin{align}
e^B & = R/2 ,  \\
F & = - J + \d\left[ e^{-B} ( \d u + P ) \right] ,  
\end{align}
where $J$ is the K\"ahler form of ${\cal M}_8$. The equation \eqref{ke} was in fact 
first derived from a study of D3-branes wrapped on K\"ahler two-cycle in \cite{Kim:2005ez}.

One can rewrite the $D=11$ solutions \eqref{11sol}
in a form compatible with \eqref{m2conf} and check the supersymmetry 
conditions given above, in particular \eqref{ke}. The symmetry of the solutions in $D=11$ imply that  the eight-dimensional base manifold
should contain factors of $H^2$ and $\mathbb{CP}^2$. Instead of just checking the BPS
conditions for our $D=11$ solutions, we here propose an ansatz suitable for a 
generalization to other dimensions \cite{Gauntlett:2007ts}. 
The point is that the same equation \eqref{ke}, when solved for six-dimensional K\"ahler space,
can be used to construct $AdS_3$ solutions in IIB supergravity from D3-branes
wrapped on 2-cycles \cite{Kim:2005ez}. And it was also shown that
there exists a generalization to arbitrary higher dimensions \cite{Gauntlett:2007ts}, with 
a gravity action coupled to a vector field, a dilaton, and an associated set of BPS equations.
In fact such a generalization was already obtained in \cite{Kim:2012ek}, for the old wrapped M2-brane $AdS_2$ solutions in \cite{Gauntlett:2001qs} and the result here presents a slight 
generalization. 

For the $AdS_2$ solutions from wrapped M2-branes 
our metric ansatz for K\"ahler space is 
\be
ds^2_{2n+4} = \Delta_1 \d s^2(H^2)  +  \Delta_2   \d \theta^2 + \frac{\sin^2 2\theta}{\Delta_2}  D\psi^2 + \cos^2\theta \d s^2(KE^+_{2n}) , 
\ee
where $KE^+_{2n}$ is $\mathbb{CP}^2$ for wrapped M2-brane solutions. In general they 
are K\"ahler-Einstein space with unit radius and satisfy 
\bea
\d \Omega_2 &=& i P_2 \wedge \Omega_2, \quad
\d P_2 = -J_2 , 
\\
\d \Omega_{2n} &=& i P_{2n} \wedge \Omega_{2n}, \quad
\d P_{2n} = J_{2n} . 
\eea
The nontrivial $U(1)$ fibration $D\psi$ in our metric ansatz is given as 
\be
D\psi = d\psi - a P_2 - P_{2n} . 
\ee
where $a$ is a constant. Recall that for the old solutions in \cite{Gauntlett:2001qs}, 
the uplifted solution has $a=1$. One may check that $z\neq 1/3$ leads to $a\neq 1$, 
after straightforward re-phrasing of our uplifted solution \eqref{11sol} into the form 
of \eqref{m2conf}.

Obviously the K\"ahler form should be
\be 
J_{2n+4} = \Delta_1 J_2 + \sin 2\theta \, \d \theta \wedge D\psi + \cos^2 \theta J_{2n}
\ee
From its closure, we easily get ($b$ is a constant.)
\be 
\Delta_1 = a \sin^2 \theta + b . 
\ee
We may calculate the Ricci tensor from the exterior derivative of $(n+2,0)$ form $\Omega$. 
\be
\Omega = e^{-i \psi} \Omega_2 \wedge \cos^n \theta 
\left(
\sqrt{\Delta_1\Delta_2} \d \theta + i \sqrt{\Delta_1/\Delta_2} \sin^2\theta D\psi \right)
\wedge \Omega_{2n}
\ee
From $\d \Omega = i P \wedge \Omega$, we have 
\be
P_{2n+4} = - \frac{1}{\cos^n\theta \sqrt{\Delta_1\Delta_2}} \frac{\d}{\d \theta}
\left[ 
\sqrt{\Delta_1/\Delta_2} \sin 2\theta \cos^n \theta
\right]
 - \d \psi + P_2 + P_{2n} . 
\ee
And it is also straightforward to compute Ricci scalar and other invariants such as $R_{ij}R^{ij}$ etc. 

Not surprisingly, for the most general case  the expression for the scalar curvature $R$ for 
K\"ahler base space is quite involved. It simplifies greatly if we assume $\Delta_2=c^2\Delta_1$, which is of course consistent with our uplifted solution. If we further demand that $R\propto (a\sin^2\theta+b)^{-1}$  as the identification of warp factor $\Delta$ as $e^{-B}$ very clearly shows, 
we need to set
$c^2 = 2(n+1)/(a+b)$. Then we have
\be
R = \frac{2(2a+(n+2)b-1)}{a \sin^2\theta + b} . 
\ee
Furthermore, if we consider the entire expression in \eqref{ke}, we find that it is factorized with 
a numerical coefficient, so the 4-th order differential equation can be reduced 
to a quadratic equation
\be
a^2 - 2n ab -b ( n(n+2) b -2(n+1) ) = 0 . 
\label{qe2}
\ee
One can easily check the old solution of M2-branes wrapped on 2-cycle in CY4 \cite{Gauntlett:2001qs}\cite{Kim:2012ek}, which corresponds to $a=1,b=1/n$. 

Although it is a good sign that we again have obtained a quadratic equation, 
it is not immediately clear that the two quadratic equations \eqref{qe1} and \eqref{qe2} in fact
encode the same $D=11$ solutions. A neat way would be 
to compare two physically significant quantities. One is
the ratio of radii as given in \eqref{sr}, and another is  the ratio of the maximal and minimal values 
of warp factor $\Delta_1$ and equivalently $\Delta$ as a function of $\theta$. By identifying these two quantities, we
find the following mapping of parameters
\be
e^{2\phi} = 1+a/b, \quad z = (a+2b)/6 . 
\ee
And then it is easy to check the equivalence of \eqref{qe1} and \eqref{qe2}.

\section{Discussion}
In this work we studied $AdS_2$ solutions in M-theory by applying the prescription of \cite{Bah:2011vv}\cite{Bah:2012dg} 
to M2-branes wrapped on 2-cycles with 
non-abelian connections.
As one varies the parameter $z$, the solutions interpolate CY4 and CY5. As a consistency 
check we have verified that the uplifted $D=11$ solutions satisfy the general supersymmetry
condition \eqref{ke} and we also presented a generalization to other dimensions. 

With new $AdS$ solutions in String/M-theory in general, it is natural to try to 
identify the dual conformal field theory and check the validity. For the solutions at hand, 
the duals are (strongly coupled) quantum mechanics. 
The problem is, unlike M5 and D3-branes wrapped on 
2-cycles, there is not yet a protected quantity such as central charges which can be 
conveniently calculated on both sides for comparison. The parameter $z$ dictates how $U(1)_R$ symmetry
should be given as a linear combination of global $U(1)$ symmetries, and  a rule such as a-maximization 
\cite{Intriligator:2003jj} or c-extremization  \cite{Benini:2012cz} is yet to be discovered. 
It would be certainly very nice
if we can establish AdS/CFT duality pairs in a quantitative manner, probably in a way similar to 
F-maximization \cite{Jafferis:2011zi} where one uses the non-divergent part of the partition function 
as a function of $R$-charges. 
We hope that the one-parameter extensions and their Killing spinor geometry 
in this paper help to uncover such relations. 

The wrapped brane constructions in gauged supergravity theories were generalized 
to 3 and 4-cycles in \cite{Acharya:2000mu}\cite{Gauntlett:2000ng}\cite{Gauntlett:2001jj}.
It would be interesting to find new solutions which interpolates different wrapped brane $AdS$ 
solutions with different amount of supersymmetries. We plan to come back to this issue
in the near future.
\begin{acknowledgments}
This work was supported by 
National Research Foundation of Korea (NRF) grants funded by the Korea government (MEST) with grant No. 2010-0023121, 2012046278, and also through the Center for Quantum Spacetime(CQUeST) of Sogang University with grant number 2005-0049409. 
\end{acknowledgments}
\bibliography{m2w}
\end{document}